\documentclass{iopjournal}

\usepackage{amsthm}

\theoremstyle{definition}
\newtheorem{definition}{Definition}[subsection]

\usepackage{amsfonts}
\usepackage{amsmath}
\usepackage{subcaption}

\begin{document}

\articletype{Paper} %	 e.g. Paper, Letter, Topical Review...

\title{Resolving Turbulent Magnetohydrodynamics: A Hybrid Operator-Diffusion Framework}

\author{Semih Kacmaz$^{1,2, *}$\orcid{0000-0002-2714-6793}, E. A. Huerta$^{2,3,4,5}$\orcid{0000-0002-9682-3604} and Roland Haas$^{1, 6}$\orcid{0000-0003-1424-6178}}

\affil{$^1$National Center for Supercomputing Applications, University of Illinois Urbana-Champaign, Urbana, Illinois 61801, USA}

\affil{$^2$Department of Physics, University of Illinois Urbana-Champaign, Urbana, Illinois 61801, USA}

\affil{$^3$Data Science and Learning Division, Argonne National Laboratory, Lemont, Illinois 60439, USA}

\affil{$^4$Department of Computer Science, The University of Chicago, Chicago, Illinois 60637, USA}

\affil{$^5$Department of Astronomy, University of Illinois Urbana-Champaign, Urbana, Illinois 61801, USA}

\affil{$^6$Department of Physics and Astronomy, University of British Columbia, Vancouver, BC V6T 1Z1, Canada}

\affil{$^*$Author to whom any correspondence should be addressed.}

\email{skacmaz2@illinois.edu}


\begin{abstract}
We present a hybrid machine learning framework that combines Physics-Informed Neural Operators (PINOs) with score-based generative diffusion models to simulate the full spatio-temporal evolution of two-dimensional, incompressible, resistive magnetohydrodynamic (MHD) turbulence across a broad range of Reynolds numbers ($\mathrm{Re}$). The framework leverages the equation-constrained generalization capabilities of PINOs to predict coherent, low-frequency dynamics, while a conditional diffusion model stochastically corrects high-frequency residuals, enabling accurate modeling of fully developed turbulence. Trained on a comprehensive ensemble of high-fidelity simulations with $\mathrm{Re} \in \{100, 250, 500, 750, 1000, 3000, 10000\}$, the approach achieves state-of-the-art accuracy in regimes previously inaccessible to deterministic surrogates. At $\mathrm{Re}=1000$ and $3000$, the model faithfully reconstructs the full spectral energy distributions of both velocity and magnetic fields late into the simulation, capturing non-Gaussian statistics, intermittent structures, and cross-field correlations with high fidelity. At extreme turbulence levels ($\mathrm{Re}=10000$), it remains the first surrogate capable of recovering the high-wavenumber evolution of the magnetic field, preserving large-scale morphology and enabling statistically meaningful predictions. 
\end{abstract}

\section{Introduction}
\label{introduction}

Magnetized fluid turbulence—characterized by intricate coupling between fluid and magnetic fields—remains one of the most formidable challenges in theoretical and computational physics. Beyond its intrinsic theoretical appeal, the study of magnetohydrodynamic (MHD) turbulence carries significant scientific and practical importance. In astrophysics, MHD turbulence underpins the transport of angular momentum in accretion disks \cite{Balbus1991,Brandenburg2005}, governs the heating of the solar corona \cite{Parker1979, Matthaeus2011}, and regulates the structure of the interstellar medium \cite{Armstrong1995, Brandenburg2013}. In geophysics, it plays a central role in the generation and sustenance of planetary dynamos, including Earth’s magnetic field \cite{Roberts2013}. In laboratory and engineering contexts, understanding MHD turbulence is crucial for optimizing magnetic confinement in fusion devices such as tokamaks and stellarators \cite{Graves_2025, Terry2000}, as well as for liquid-metal cooling technologies in next-generation nuclear reactors \cite{Molokov2007}. Across these systems, nonlinear couplings between velocity and magnetic fields govern cascade processes, intermittency, and reconnection events that directly impact energy transport, stability, and confinement. Thus, accurate modeling of MHD turbulence is indispensable not only for advancing fundamental plasma physics but also for enabling predictive capability in astrophysical, geophysical, and technological applications.

The two-dimensional, incompressible, resistive magnetofluid, defined by the magnetohydrodynamic (MHD) equations, hosts a rich tapestry of phenomena: Lorentz-force–driven advection, resistive diffusion, current-sheet formation, magnetic reconnection, and energetic cascades spanning scales. Even in this simplified two-dimensional, incompressible regime, the dynamics far exceed the complexity of traditional testbeds such as Kolmogorov flow, 2D Navier–Stokes turbulence in vorticity form, steady-state Darcy flow, Burgers equations in the scalar, inviscid and vector types, and the 2D linear and nonlinear shallow water equations~\cite{li2024physics,Rosofsky:2022lgb}.

Kolmogorov flow—born from sinusoidal forcing in periodic domains—is a classical setting for energy cascade and large-scale structure formation in a single velocity field. However, it lacks the magnetic coupling and Lorentz feedback that define MHD. The 2D Navier–Stokes system in vorticity formulation introduces a single advective nonlinearity and is regulated solely by viscosity, enabling chaotic but comparatively tractable dynamics. Steady-state Darcy flow, in contrast, is linear, elliptic, and static, devoid of temporal evolution or feedback mechanisms—its simplicity is hardly commensurate with the demands of turbulent MHD.

Resistive 2D MHD, by contrast, comprises two strongly coupled nonlinear partial differential equations: one governing vorticity, influenced by both advective transport and Lorentz forces, and another for magnetic induction, shaped by advection and resistive decay. Each field obeys its own cascade and dissipation scales, and the nonlinear feedback between them spawns small-scale intermittent structures, spectral non-Gaussianity, and spontaneous reconnection events—phenomena that surge in intensity at Reynolds numbers (Re) near 500, straining conventional numerical solvers and machine learning surrogates.

Machine learning approaches grounded in physics have sought to tame turbulence, but each has exhibited inherent limitations. Physics-Informed Neural Networks (PINNs) embed PDE constraints via pointwise residual minimization. They have proven effective in laminar and simple turbulent regimes—e.g., Reynolds-Averaged Navier-Stokes (RANS) and boundary-layer flows—but struggle with stiffness, high $\mathrm{Re}$, and complex multiscale coupling, often due to ill-conditioned optimization landscapes and spectral mismatches~\cite{Raissi2019PIDD}.

Physics-Informed Neural Operators (PINOs) address some of these deficits by learning mappings between function spaces\cite{Li2021PINO, LuDeepONet2021, KovachkiNeurops2023}. Pioneered by Li \textit{et al.}~\cite{li2024physics}, PINOs extend the Fourier Neural Operator (FNO) with physics-based losses, enabling data-efficient, super-resolution capabilities and strong performance on canonical PDEs like Burgers’, Darcy, and Navier–Stokes. Rosofsky \textit{et al.} applied PINOs to 2D incompressible MHD using tensor-FNO, achieving high accuracy up to $\mathrm{Re} \approx 250$ across grid resolutions up to $256^2$~\cite{Rosofsky2023MHD}. However, their study revealed increasing mean-squared errors in small-scale magnetic structures as Re exceeded 500, underscoring the limitations of deterministic operator surrogates in turbulent regimes.

Generative diffusion models offer a powerful complement. These models, such as Denoising Diffusion Probabilistic Models (DDPMs) and score-based diffusion frameworks, learn complex distributions through iterative denoising and inversion of stochastic corruption processes~\cite{Ho2020DDPM, Song2021Score}. Recent works have shown the efficacy of diffusion-based surrogates in reconstructing high-fidelity turbulent flow fields from coarse measurements~\cite{Shu2022PID, LiSynthTurbulence2024} when guided by physics-informed conditioning. This concept has been generalized to solve both forward and inverse PDE problems even under extreme data sparsity, where diffusion models learn the joint distribution of physical coefficients and solutions to fill in missing information~\cite{HuangDiffusionPDE2024}. However, a fundamental challenge arises when applying these models to temporal dynamics. As identified by Guo et al.~\cite{GuoDydiff2025}, standard diffusion frameworks treat prediction as a conditional generation task and do not explicitly model the evolution between temporally adjacent latent states, potentially limiting their ability to generate long, coherent sequences. This challenge is further compounded in the MHD context, where a monolithic diffusion model would be tasked with learning the entire joint distribution of the velocity and magnetic fields. Given the complex and often subtle coupling between the fields, such an approach would be suboptimal as it could struggle to capture the crucial cross-field correlations that govern the system's evolution, instead learning the marginal distributions of each field independently.

The respective limitations of neural operators and diffusion models suggest a natural synergy. This motivates the adoption of a hybrid framework, such as the one proposed by Oommen et al.~\cite{OommenEtAl2025}, which synergistically fuses the strengths of both paradigms. In this two-stage approach, a deterministic neural operator provides the foundational spatio-temporal structure and enforces physical consistency, while a stochastic, score-based diffusion model acts as a corrector, generating the missing high-frequency turbulent details. While this methodology has shown promise, its application to a physically rich and challenging system like magnetized turbulence remains unexplored. Through this work, we harness the PINO's operator-level structure and the diffusion model's fine-scale generation capabilities—arguably reaching the necessary nuance to capture full-field statistics, coherent structures, and non-Gaussian tails in turbulence up to $\mathrm{Re} \approx 3{,}000$.

Here, we present the first PINO–Diffusion hybrid tailored for turbulent MHD. We demonstrate that it achieves unprecedented fidelity: matching spectral distributions, recovering intermittent phenomena, and significantly outperforming deterministic surrogate models. Our method not only advances the frontier of AI-driven turbulence modeling but lays the foundation for broader applications in more complex, physically rich MHD systems.

\section{Background and Methodology}
\label{sec:methodology}

This section outlines the theoretical foundations of the neural operator and score-based diffusion models. It further describes how these components were implemented and integrated for the specific physical problem addressed in this work.

\subsection{Neural Operators and Physics-Informed Training}
\label{app:fno_and_loss}

A neural operator is a generalization of a standard neural network designed to learn mappings between infinite-dimensional function spaces. Given an input function $a \in \mathcal{A}$ and a target function $u \in \mathcal{U}$, where $\mathcal{A}$ and $\mathcal{U}$ are Banach spaces, the goal is to learn the operator $G^\dagger: \mathcal{A} \to \mathcal{U}$ such that $u(x) = G^\dagger(a(x))$. The architecture approximates this operator through an iterative structure.

\begin{definition}[Neural Operator]
\label{def:neural_operator}
An operator $G_\theta: \mathcal{A} \to \mathcal{U}$ is defined by the composition $G_\theta = Q \circ (L_T \circ \dots \circ L_1) \circ P$, where $P$ is a lifting operator, $Q$ is a projection operator, and each layer $L_t$ for $t=1, \dots, T$ updates a latent function sequence $v_t \in \mathcal{V}_t$ as follows:
\begin{equation}
    v_{t+1}(x) = \sigma \left( W v_t(x) + (\mathcal{K}(a; \phi)v_t)(x) \right),
\end{equation}
where $W: \mathcal{V}_t \to \mathcal{V}_{t+1}$ is a local linear transformation, $\sigma$ is a pointwise non-linear activation function, and $\mathcal{K}$ is a kernel integral operator.
\end{definition}

The non-local interactions are captured by the kernel integral operator, which is learned from data.

\begin{definition}[Kernel Integral Operator]
\label{def:kernel_operator}
The kernel integral operator $\mathcal{K}$ parameterized by $\phi \in \Theta_\kappa$ is defined as:
\begin{equation}
    (\mathcal{K}(a; \phi)v_t)(x) = \int_{D} \kappa_\phi(x, y, a(x), a(y)) v_t(y) dy,
\end{equation}
for $x, y \in D$, where $D$ is the domain of the functions. The kernel $\kappa_\phi$ is typically parameterized by a neural network.
\end{definition}

The Fourier Neural Operator (FNO)~\cite{Li2021PINO} provides an efficient and expressive implementation of this architecture by constraining the kernel integral to be a convolution, where the kernel depends only on the displacement, $\kappa_\phi(x, y) = \kappa_\phi(x-y)$. By the convolution theorem, this operation is equivalent to a pointwise multiplication in the Fourier domain. This leads to the formal definition of the Fourier integral operator.

\begin{definition}[Fourier Integral Operator]
\label{def:fno_operator}
The Fourier integral operator $\mathcal{K}_\phi$ is defined as:
\begin{equation}
    (\mathcal{K}_\phi v_t)(x) = \mathcal{F}^{-1} \left( R_\phi \cdot (\mathcal{F}v_t) \right)(x),
\end{equation}
where $\mathcal{F}$ is the Fourier transform, $\mathcal{F}^{-1}$ is its inverse, and $R_\phi$ is the Fourier transform of the convolutional kernel $\kappa_\phi$, parameterized by the learnable weights $\phi$.
\end{definition}

In practice, $R_\phi$ is directly parameterized in the frequency domain as a complex-valued tensor. The operation is restricted to a finite set of low-frequency modes, $|\mathbf{k}| \le k_{max}$, with higher-frequency modes truncated. The use of the Fast Fourier Transform (FFT) makes this layer computationally efficient, with complexity $\mathcal{O}(N \log N)$ for a grid of size $N$, while also making the learned weights $R_\phi$ independent of the grid resolution.

To ensure physical consistency and data fidelity, the operator is trained using a composite physics-informed loss function. The total loss, $\mathcal{L}(\theta)$, for a set of model parameters $\theta$ is a weighted sum of several components:
\begin{equation}
    \mathcal{L}(\theta) = \lambda_{data} \mathcal{L}_{data} + \lambda_{ic} \mathcal{L}_{ic} + \lambda_{pde} \mathcal{L}_{pde}.
\end{equation}
The Data Loss ($\mathcal{L}_{data}$) is a standard supervised term measuring the discrepancy between the model's prediction and the ground truth. The Initial Condition Loss ($\mathcal{L}_{ic}$) specifically enforces this agreement at $t=0$. The crucial PDE Residual Loss ($\mathcal{L}_{pde}$) embeds the governing physics by penalizing the norm of the residual when the model's output is passed through the known differential operators (Eqs.~\ref{eq:navier_stokes}-\ref{eq:vector_potential}). This composite objective, inspired by Physics-Informed Neural Networks (PINNs)~\cite{Raissi2019PIDD}, guides the optimization towards solutions that are both accurate and physically consistent.

\subsection{Score-Based Diffusion Models and Conditional Generation}
\label{app:diffusion_models}

The second stage of our framework acts as a corrector: a score-based diffusion model conditioned on the low-frequency output of the trained PINO. This generative component enhances the surrogate solution by recovering fine-scale structures omitted in the operator's smooth approximation. 

This class of models is built on the principle of reversing a continuous-time stochastic differential equation (SDE) that transforms a data distribution, $p_{data}(x)$, into a simple prior distribution~\cite{Song2021Score}. A forward SDE gradually perturbs a data sample $x_0$ into a noise vector $x_T$:
\begin{equation}
    dx = f(x, t)dt + g(t)dw.
\end{equation}
Crucially, this process has a corresponding reverse-time SDE, a classical result in stochastic processes~\cite{Anderson1982ReverseTime}, which allows for generating data from noise by evolving backwards from a sample of the prior distribution:
\begin{equation}
    dx = [f(x, t) - g(t)^2 \nabla_x \log p_t(x_t)]dt + g(t)d\bar{w},
    \label{eq:reverse_sde_unconditional}
\end{equation}
where $\nabla_x \log p_t(x_t)$ is the score function of the perturbed data distribution $p_t$ at time $t$. The central task is to learn this score function with a time-dependent neural network, $s_\theta(x_t, t)$.

The framework adopted in this study extends this to conditional generation. Our goal is not to sample from the unconditional distribution $p(x_0)$, but from the conditional distribution $p(x_0|y)$, where the condition $y$ is the output of the PINO. The score of this conditional distribution can be decomposed via Bayes' rule:
\begin{equation}
    \nabla_x \log p_t(x_t|y) = \nabla_x \log p_t(x_t) + \nabla_x \log p_t(y|x_t).
    \label{eq:bayes_score}
\end{equation}
The first term is the unconditional score, and the second term, $\nabla_x \log p_t(y|x_t)$, is a guidance term that pushes the solution towards states $x_t$ that are more likely to have produced the condition $y$.

There are two primary strategies to implement conditional generation with diffusion models. One approach is to train an unconditional score model $s_\theta(x_t, t) \approx \nabla_x \log p_t(x_t)$, and then, during inference, modify the reverse diffusion process by incorporating a guidance term derived from an external classifier. This technique, known as classifier guidance, involves using the gradient of the log-probability from a separately trained classifier $\nabla_x \log p(y | x_t)$ to steer the sampling process toward samples that are more likely to belong to a desired class $y$. Classifier guidance was a key innovation for improving the fidelity and class-conditional accuracy of generated samples~\cite{Dhariwal2021Diffusion}.

However, the method used in our work, which is common in modern diffusion model implementations, takes a more direct approach that builds upon the principles of Denoising Diffusion Probabilistic Models (DDPMs)~\cite{Ho2020DDPM, ho2021classifierfree}. We train a single neural network to directly approximate the full conditional score, $s_\theta(x_t, y, t) \approx \nabla_x \log p_t(x_t|y)$. The conditioning information $y$ (the PINO output) is provided to the network as an additional input, typically by concatenating the condition $y$ with the noisy state $x_t$ along the channel dimension. The network is then trained to denoise this composite input.

By learning this direct mapping, the neural network implicitly learns to model both the unconditional score and the guidance term simultaneously. The training objective remains a form of denoising score matching, but on the conditional distribution:
\begin{equation}
    \mathcal{L}(\theta) = \mathbb{E}_{t, x_0, y, \epsilon} \left[ \lambda(\sigma_t) \| D_\theta([x_t, y]; \sigma_t) - x_0 \|^2 \right],
\end{equation}
where $x_t$ is the noised version of $x_0$, and $y$ is the corresponding condition. This approach effectively trains the denoiser $D_\theta$ to reverse the diffusion process while ensuring the final output is consistent with the provided physical prior from the PINO.

It is worth emphasizing that the efficacy of the entire framework hinges on a crucial division of labor. The score-based denoising process is inherently more effective at refining high-wavenumber components than low-wavenumber ones, where the score function gradient is weaker. Consequently, for the framework to succeed, the neural operator must provide an accurate low-frequency prior, or ‘mean flow’. A detailed analysis supporting this frequency-dependent behavior of the score function is presented in the appendix of Oommen et al.~\cite{OommenEtAl2025}.

\subsection{Governing Equations and Data Generation}
\label{subsec:data_generation}

Our study is predicated on a comprehensive dataset of high-fidelity numerical simulations, which serve as the ground truth for training and validating our surrogate models. We model the dynamics of a two-dimensional, incompressible, resistive magnetofluid governed by the standard MHD equations~\cite{Biskamp1993MHD}:
\begin{equation}
    \frac{\partial \mathbf{u}}{\partial t} + (\mathbf{u} \cdot \nabla)\mathbf{u} = -\frac{1}{\rho_0}\nabla p + \frac{1}{\rho_0}(\nabla \times \mathbf{B}) \times \mathbf{B} + \nu \nabla^2 \mathbf{u},
    \label{eq:navier_stokes}
\end{equation}
\begin{equation}
    \frac{\partial \mathbf{B}}{\partial t} = \nabla \times (\mathbf{u} \times \mathbf{B}) + \eta \nabla^2 \mathbf{B},
    \label{eq:induction}
\end{equation}
subject to the constraints $\nabla \cdot \mathbf{u} = 0$ and $\nabla \cdot \mathbf{B} = 0$. Here, $\mathbf{u}$ is the fluid velocity, $\mathbf{B}$ is the magnetic field, $p$ is the pressure, $\rho_0=1$ is the constant fluid density, $\nu$ is the kinematic viscosity, and $\eta$ is the magnetic diffusivity. The kinetic and magnetic Reynolds numbers are defined as $\mathrm{Re} = 1/\nu$ and $\mathrm{Re_m} = 1/\eta$, respectively. For this work, we consider a magnetic Prandtl number $\mathrm{Pr_m} = \nu/\eta = 1$, such that $\mathrm{Re} = \mathrm{Re_m}$. We deliberately restrict our attention to the two-dimensional, incompressible, resistive MHD equations at a unit magnetic Prandtl number. This controlled regime provides a stringent yet interpretable testbed, retaining essential turbulent features such as nonlinear cascades and magnetic reconnection, while allowing for reproducible dataset generation and systematic validation of physics-informed AI surrogates. We note, however, that the realm of applicability of the present results does not extend to fully three-dimensional turbulence, where anisotropy, Alfvénic dynamics, and small-scale intermittency introduce qualitatively new challenges. Likewise, regimes with magnetic Prandtl numbers far from unity---which are relevant for both astrophysical and laboratory plasmas---exhibit distinct dissipative scale separations and dynamo behaviors that are not addressed here. These extensions require substantially more work for dataset production and careful hyperparameter tuning for stable training, and are the subject of ongoing work. The aim of this manuscript is therefore focused: to establish, within this well-defined 2D Prandtl number $\mathrm{Pr_m} = 1$ setting, that turbulent magnetized flows can be effectively modeled using physics-informed neural operators augmented by diffusion-based generative corrections.

To satisfy the solenoidal constraint on the magnetic field to machine precision, we evolve the magnetic vector potential $\mathbf{A}$ (where $\mathbf{B} = \nabla \times \mathbf{A}$), whose evolution is governed by:
\begin{equation}
    \frac{\partial \mathbf{A}}{\partial t} + (\mathbf{u} \cdot \nabla)\mathbf{A} = \eta \nabla^2 \mathbf{A}.
    \label{eq:vector_potential}
\end{equation}

This system of equations was solved numerically using the open-source spectral solver \texttt{Dedalus}~\cite{Dedalus2020}, closely following the numerical framework established in Ref.~\cite{Rosofsky2023MHD}. The simulations were configured on a doubly periodic 2D domain $[0, 1]^2$, which was discretized using a Fourier basis with a primary resolution of $128 \times 128$ grid points and a de-aliasing factor of $3/2$. A fourth-order Runge-Kutta scheme was employed for time integration with a fixed timestep of $\Delta t = 10^{-3}$, evolving the system to a final time of $t=1.0$ with snapshots saved every $0.01$ time units. This process yields 101 high-resolution temporal frames for each simulation, which were then temporally subsampled by a factor of 4 to produce the final 26-frame sequences used for training and evaluation. To create a diverse ensemble of scenarios, initial conditions were generated from Gaussian Random Fields (GRFs)~\cite{RasmussenGRF2006} with a characteristic length scale of $l=0.1$, ensuring the initial velocity and magnetic fields were divergence-free. To investigate the transition from laminar to turbulent flow, we generated distinct datasets for a wide range of Reynolds numbers, specifically $\mathrm{Re} \in \{100, 250, 500, 750, 1000, 3000, 10000\}$. The generation of this extensive dataset was performed on the NCSA Delta cluster, where we employed a massively parallel approach to distribute thousands of independent simulation instances using all 128 CPU cores of a compute node. For each Reynolds number, this process yielded an ensemble of 1000 simulations, which was randomly partitioned into a training set of 800 samples, a validation set of 100 samples, and a test set of 100 samples.

\subsection{Diffusion-Enhanced Operator Framework}
\label{subsec:framework}

\begin{figure*}[t!]
    \centering
    \includegraphics[width=\textwidth]{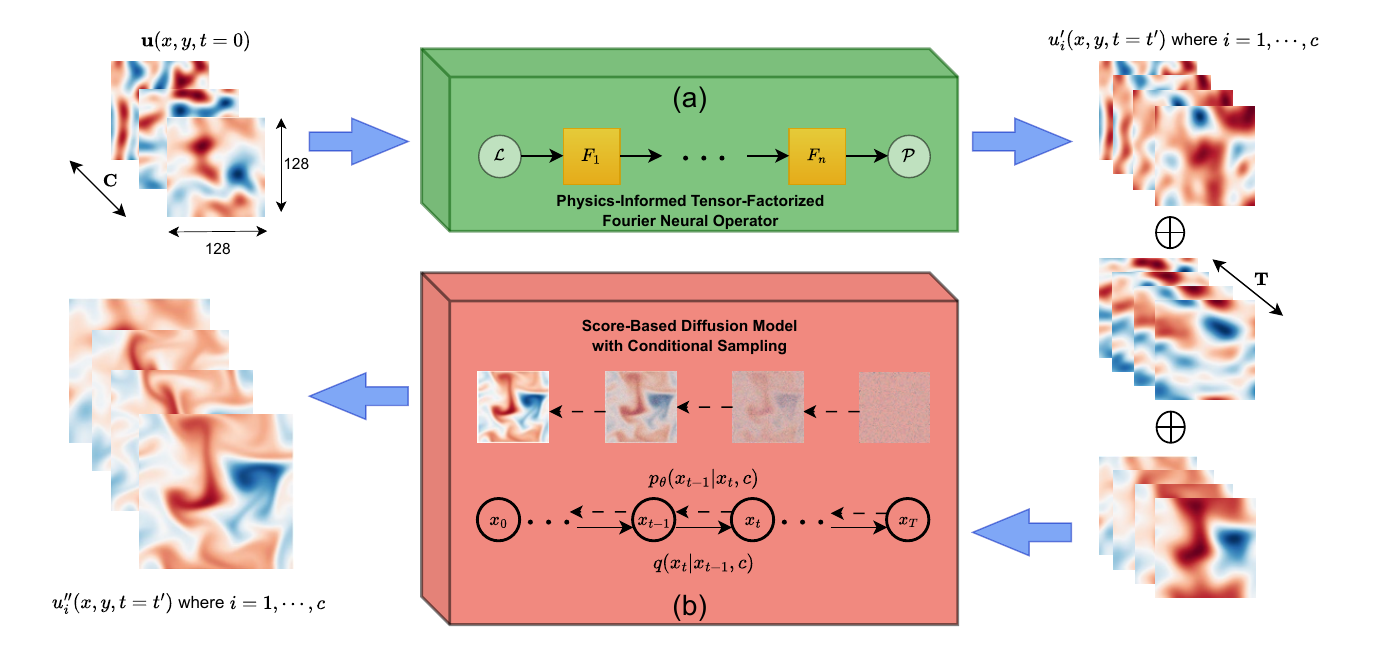}
    \caption{Schematic of the two-stage diffusion-enhanced operator framework. The process begins with the initial state of the MHD system, a three-channel stack representing $\mathbf{u}(x,y,t=0)$ at a $128 \times 128$ resolution. This state is fed into (a) the Physics-Informed Tensor-Factorized Fourier Neural Operator (PINO), which contains a lifting network ($\mathcal{L}$), a series of Fourier layers ($F_i$), and a projection network ($\mathcal{P}$). The PINO predicts the full spatio-temporal evolution, $\mathbf{u}'(x,y,t)$. These temporal sequences for each channel are then flattened and concatenated (indicated by the $\oplus$ operators) to serve as the conditional prior for the second stage. This prior is then input to (b) the Score-Based Diffusion Model, which, as conceptually illustrated by its reverse denoising process, stochastically refines the smooth PINO prediction. The final output of the pipeline, $\mathbf{u}''$, is a high-fidelity, spectrally-corrected prediction of the system's evolution.}
\label{fig:framework}
\end{figure*}

To address the spectral limitations inherent in standard neural operators when modeling turbulent regimes, we implement a two-stage, diffusion-enhanced framework, illustrated schematically in Fig.~\ref{fig:framework}. This approach, inspired by recent advances in generative modeling~\cite{OommenEtAl2025}, synergistically combines a deterministic operator for low-frequency dynamics with a stochastic model for high-frequency details.

The first stage of our pipeline employs a Physics-Informed Tensor-Factorized Fourier Neural Operator (PINO)~\cite{Li2021PINO, li2024physics} to learn the mean-field evolution of the MHD system. The architecture, an extension of the model used in Ref.~\cite{Rosofsky2023MHD}, is designed to process 3D spatio-temporal data ($x, y, t$). It consists of 8 Fourier layers with 32 latent channels. Within each spectral convolution, we retain 8 Fourier modes and utilize a Canonical Polyadic (CP) tensor decomposition with a rank of 0.5 to factorize the spectral weights~\cite{KossaifiEtAl, Kolda2009}. This factorization enhances model expressivity while maintaining computational tractability on large-scale hardware. Crucially, the PINO is trained with a composite loss function that, in addition to a standard data-fitting term, includes residuals of the governing MHD equations (Eqs.~\ref{eq:navier_stokes}-\ref{eq:vector_potential}) as soft constraints. The PINO was trained on a single AMD MI250x GPU on the Frontier supercomputer at the Oak Ridge Leadership Computing Facility (OLCF).

The output of the trained PINO, which represents a smooth, spectrally-limited approximation of the true solution, serves as a conditional prior for the second stage: a score-based diffusion model~\cite{Song2021Score}. This generative model is built upon a UNet backbone~\cite{RonnebergerUNet2015} with a base dimension of 128 and six down/up-sampling stages, defined by channel multipliers of (1, 2, 3, 5, 8, 12). Self-attention, accelerated via Flash Attention~\cite{Dao2022flashattention, Dao2023flashattention2}, is incorporated at each resolution with 8 heads and a head dimension of 64. The network operates within a preconditioned diffusion framework that uses calibrated noise schedules and input normalizations to balance the learning dynamics. This design has been shown to enhance training efficiency and sample fidelity in recent diffusion models~\cite{KarrasEdm2022}. The model is tasked with stochastically generating the high-wavenumber turbulent structures absent from the PINO's prediction. During its reverse-denoising process, it is explicitly conditioned on the PINO output, ensuring that the generated turbulent structures are physically consistent with the underlying large-scale flow. The training data for the diffusion model, consisting of pairs of PINO predictions and their corresponding ground truth simulation states, was generated on Frontier, while the model itself was trained on NVIDIA H100 GPUs on the NCSA Delta AI cluster.

In inference, this pipelined approach first generates a fast, low-frequency prediction with the PINO, which is then refined by the conditional diffusion model to yield a final, high-fidelity state that accurately captures the full energy spectrum of the turbulent flow.

\section{Results and Discussion}
\label{sec:results}

We present a comprehensive analysis of our diffusion-enhanced framework. We begin by establishing the performance of PINOs as powerful yet inherently limited baselines. We then demonstrate how the integration of a conditional diffusion model rectifies these limitations, providing robust quantitative and qualitative evidence of the combined framework's superior performance across a wide range of Reynolds numbers.

\subsection{PINO Performance and Inherent Limitations}
\label{subsec:pino_limitations}

\begin{figure*}[ht!]
    \centering
    \includegraphics[width=\textwidth]{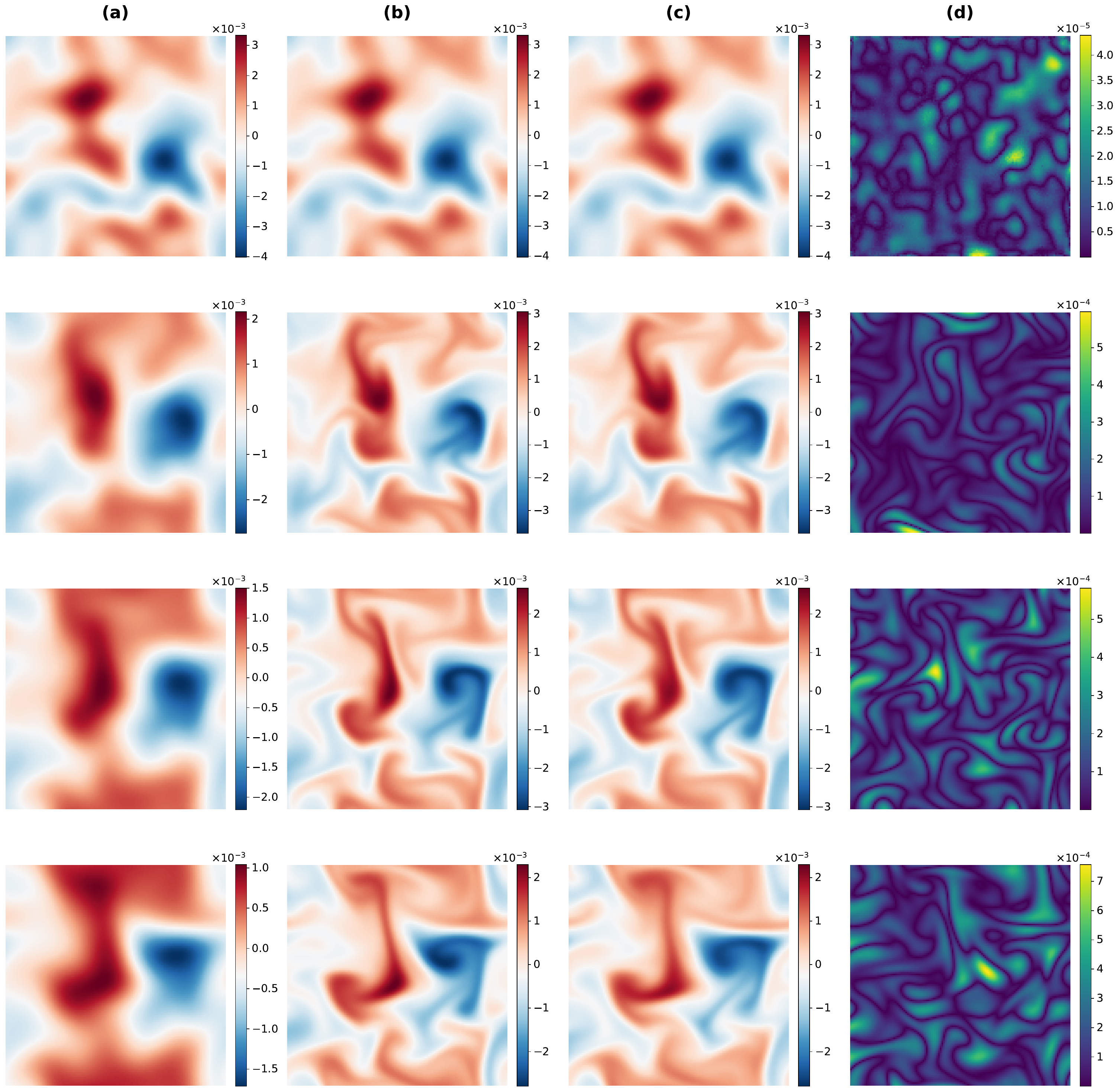}
    \caption{Temporal evolution of the vector potential ($A$) for Reynolds-number $\mathrm{Re}=1000$. The rows display the system at four distinct time steps, corresponding to $t=0, 0.33, 0.67,$ and $1.0$ from top to bottom. Columns show: (a) the smooth PINO-only prediction, which acts as the conditional prior, (b) the ground truth, (c) the final high-fidelity prediction from the PINO+Diffusion framework, and (d) the absolute error of the final prediction. The framework consistently refines the smooth prior into an accurate, detailed state.}
    \label{fig:diffusion_re1000}
\end{figure*}

Our PINO model, featuring a larger architecture (8 Fourier layers, 32 latent channels) than that used in prior work~\cite{Rosofsky2023MHD}, serves as a powerful baseline for this study. This increased capacity allows the model to accurately predict the full spatio-temporal evolution for flows in the laminar and early transitional regimes. However, as the Reynolds number increases, the inherent architectural limitations of the FNO core, upon which our PINO is built, become the dominant factor in performance.

The FNO~\cite{Li2021PINO} architecture operates by transforming the input into Fourier space and applying a learned transformation to a fixed, predefined number of low-frequency modes. This operation acts as an intrinsic low-pass filter, as any information residing in wavenumbers higher than the truncation cutoff is explicitly discarded by the spectral convolution layers. While effective for smooth functions, this design is structurally incapable of representing the high-frequency information that characterizes more complex flows.

This architectural limitation becomes pronounced in high-Reynolds-number flows. This is illustrated by the PINO-only predictions for the magnetic vector potential ($A$) at $\mathrm{Re}=1000$, shown in column (a) of Fig.~\ref{fig:diffusion_re1000}. The model's failure is most acute for the magnetic vector potential channel, as its predictions become progressively and significantly smoothed while the large-scale velocity fields remain comparatively better resolved. This qualitatively demonstrates that as more energy shifts to higher wavenumbers in higher-Re flows, the PINO's spectral bias prevents it from accurately modeling the system, motivating the need for our diffusion-based enhancement.

\begin{figure*}[ht!]
    \centering
    \includegraphics[width=\textwidth]{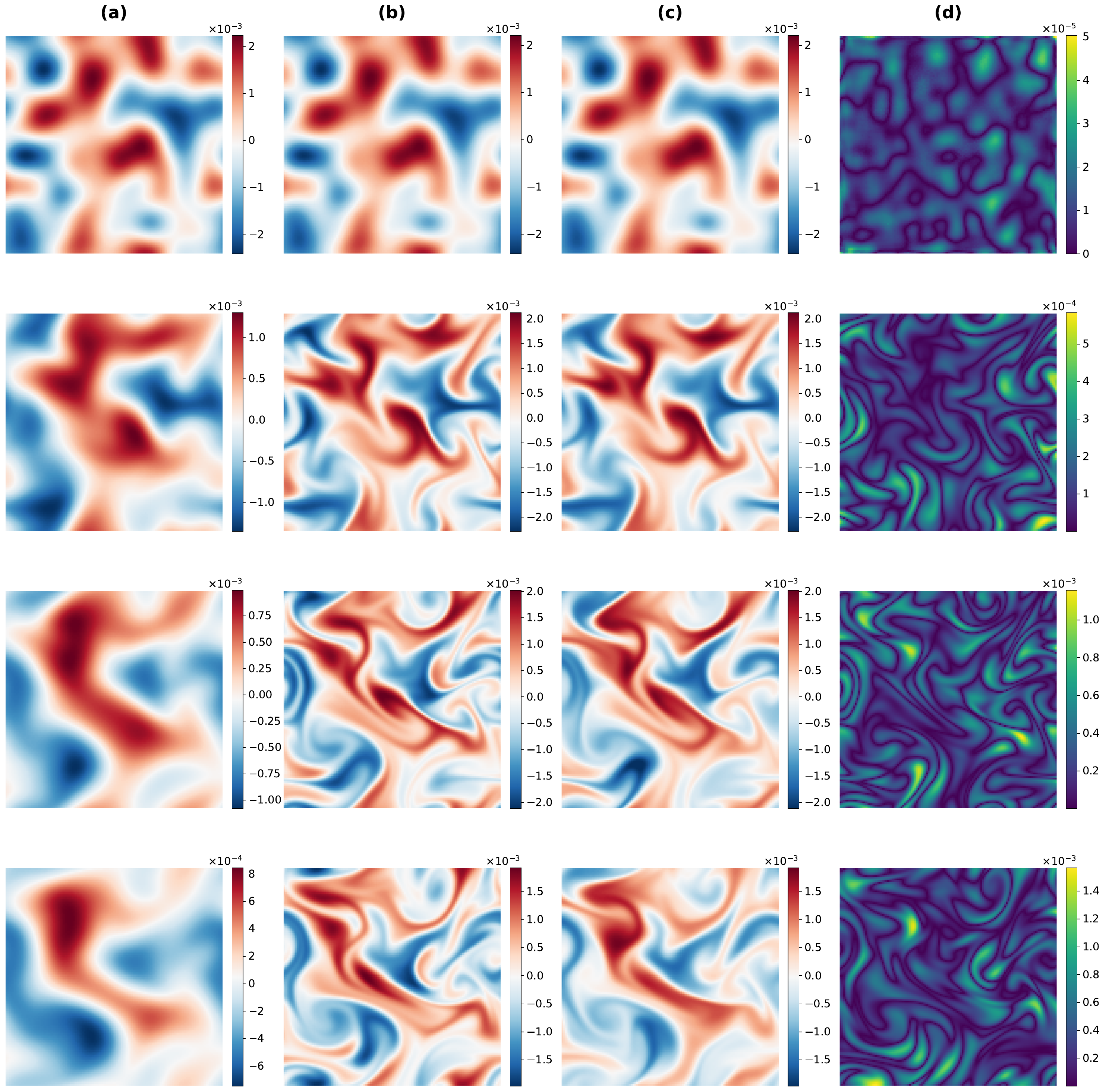}
    \caption{Temporal evolution of the vector potential ($A$) for a turbulent case at $\mathrm{Re}=3000$. The layout is identical to Fig.~\ref{fig:diffusion_re1000}. The framework continues to successfully refine the smooth PINO prior (a) into a high-fidelity prediction (c) that closely matches the ground truth (b).}
    \label{fig:diffusion_re3000}
\end{figure*}

\subsection{Spectral Correction via Diffusion-Enhanced Framework}
\label{subsec:spectral_correction}

\begin{figure*}[ht!]
    \centering
    \begin{subfigure}[b]{0.49\textwidth}
        \centering
        \centerline{\normalsize (a) $\mathrm{Re} = 1000$}
        \vspace{1mm}
        \includegraphics[width=\textwidth]{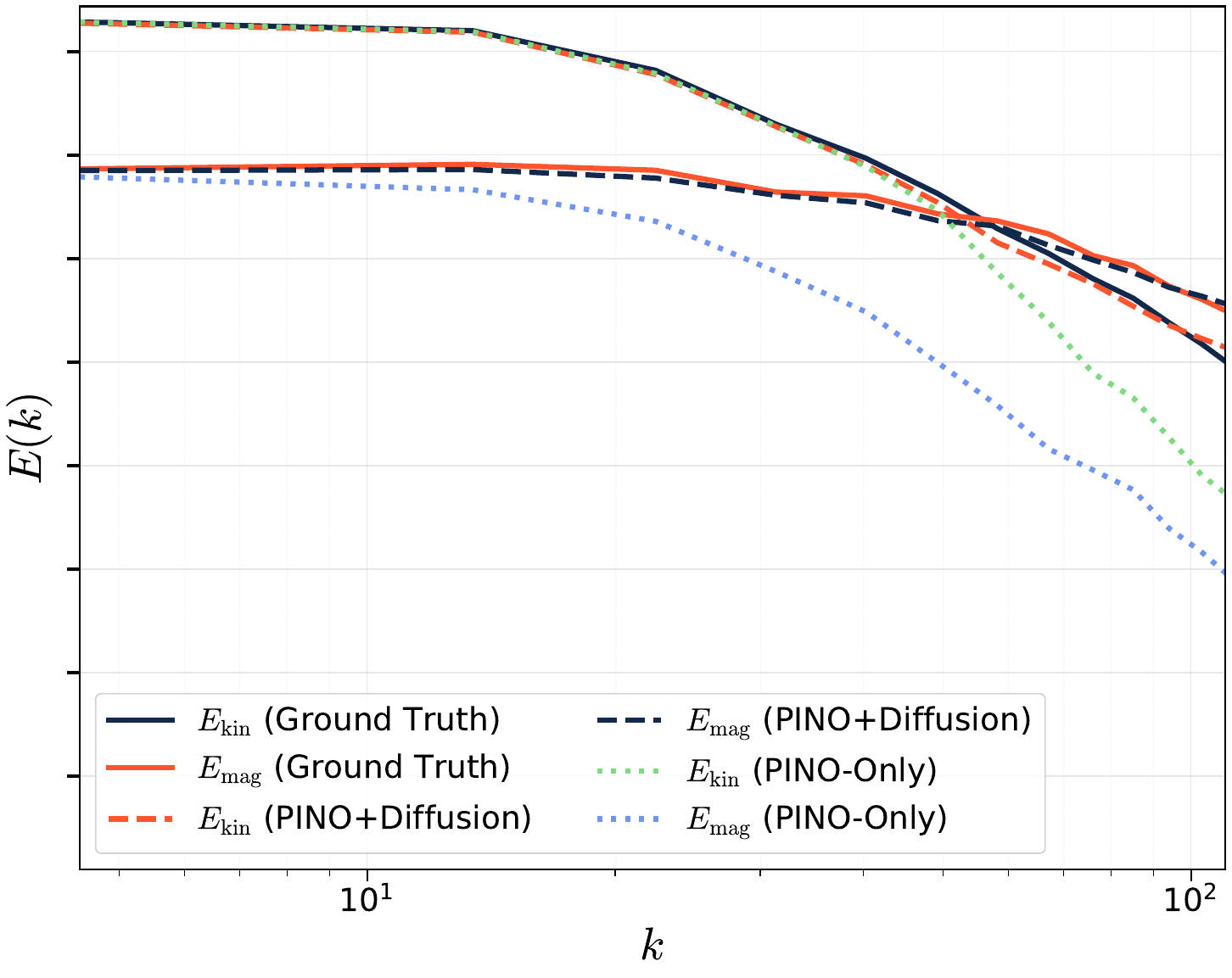}
        % \label{fig:spectra_re1000_sub}
    \end{subfigure}% <--- The '%' is crucial. It removes the unwanted space.
    \hfill 
    \begin{subfigure}[b]{0.49\textwidth}
        \centering
        \centerline{\normalsize (b) $\mathrm{Re} = 3000$}
        \vspace{1mm}
        \includegraphics[width=\textwidth]{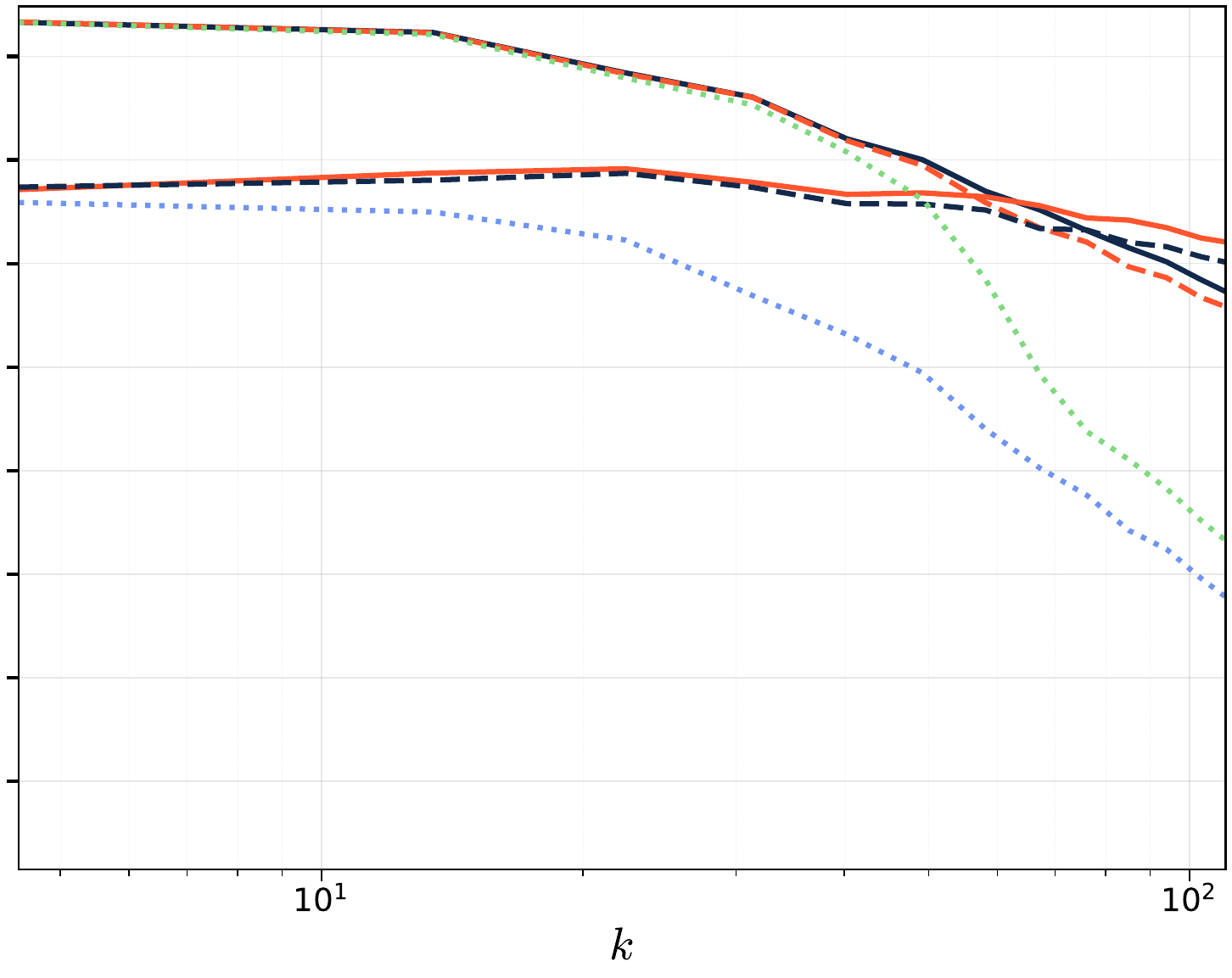}
        % \label{fig:spectra_re3000_sub}
    \end{subfigure}
    \caption{Kinetic and magnetic energy spectra, $(E_{\mathrm{kin}},E_{\mathrm{mag}})$, at $t=1.0$ for representative cases at (a) $\mathrm{Re}=1000$ (left panel) and (b) $\mathrm{Re}=3000$ (right panel). In both regimes, the PINO+Diffusion model (dashed lines) accurately reconstructs the ground truth spectra (solid lines) across all wavenumbers. In contrast, the PINO-only model (dotted lines) exhibits a significant spectral bias, failing to capture high-wavenumber energy. This failure is consistently more pronounced for the magnetic field, quantitatively demonstrating the efficacy of the diffusion-based correction.}
    \label{fig:spectra_comparison}
\end{figure*}

To address the spectral bias identified in the PINO, the framework adopted in this study employs a two-stage, diffusion-enhanced pipeline. This approach represents a strategic division of the learning task, motivated by fundamental challenges in applying generative models to temporal PDE dynamics. Standard diffusion models often treat prediction as a simple conditional generation task, failing to explicitly model the transitions between temporally adjacent latent states and thus struggling to produce temporally coherent sequences~\cite{GuoDydiff2025}. Furthermore, while diffusion models are well-suited to handle uncertainty from partial observations by learning a joint distribution over the system state~\cite{HuangDiffusionPDE2024}, tasking a single model with learning both the deterministic bulk evolution and the high-frequency stochastic details is a formidable challenge. The framework circumvents this by using the PINO to efficiently compute a deterministic, low-frequency prior, leaving the diffusion model with the more targeted task of acting as a stochastic corrector for the high-frequency residual.

The generative capability of this corrector stage arises from training via denoising score matching~\cite{VincentScoreMatching2011}, where the model learns to approximate the score function, $\nabla_x \log p(x; \sigma)$, across a range of noise scales $\sigma$. Inference proceeds by numerically solving a reverse-time stochastic differential equation (SDE), gradually transforming an initial noise sample into a realistic output by following the learned score field. The addition of controlled stochasticity at each step mitigates error accumulation and enhances sample quality~\cite{KarrasEdm2022}.

The efficacy of this complete framework is demonstrated in Fig.~\ref{fig:diffusion_re1000} for a high-Reynolds-number case at $\mathrm{Re}=1000$. Using the smoothed PINO-only prediction (column a) as a conditional prior, the diffusion model refines the solution. The resulting prediction from the full framework (column c) visually matches the ground truth (column b) with high fidelity at all depicted time steps. The successful reconstruction of the flow's fine details is confirmed by the substantially reduced error, as shown in column (d).

We have also quantified the improvement of our PINO+Diffusion framework by analyzing the energy distribution in Fourier space. Fig.~\ref{fig:spectra_comparison} presents the kinetic and magnetic energy spectra at $t=1.0$ for $\mathrm{Re}=\{1000, 3000\}$, the final simulation time chosen to evaluate the models where cumulative error is most significant. The PINO+Diffusion approach (dashed lines) accurately reconstructs the ground truth spectra (solid lines) for both kinetic and magnetic energy across all wavenumbers. In stark contrast, the PINO-only model (dotted lines) exhibits a significant spectral bias, failing to capture high-wavenumber energy. This failure is more pronounced for the magnetic field, whose spectrum deviates at a lower wavenumber, quantitatively demonstrating the efficacy of the diffusion-based correction.

\subsection{Quantitative Analysis Across Reynolds Numbers}

To provide a comprehensive quantitative assessment, we report the time-averaged relative L2 error on the unseen test set for both models across all tested Reynolds numbers. The results are summarized in Table~\ref{tab:l2_errors}. It is important to note that these errors are computed on data channels that have been independently normalized to the range [-1, 1]. This ensures that each physical field contributes equitably to the total error, despite the inherent scale differences between the velocity and magnetic vector potential.

Formally, for a predicted spatio-temporal field $\mathbf{u}_{\text{pred}}$ and a ground truth field $\mathbf{u}_{\text{true}}$, we calculate the relative L2 error $\mathcal{E}$ on the test set by averaging over the $N_{\text{test}}$ samples as follows:

\begin{equation}
\mathcal{E} = \frac{1}{N_{\text{test}}} \sum_{i=1}^{N_{\text{test}}} \frac{\|\mathbf{u}_{\text{pred}}^{(i)} - \mathbf{u}_{\text{true}}^{(i)}\|_{2}}{\|\mathbf{u}_{\text{true}}^{(i)}\|_{2}},
\label{eq:rel_l2_error}
\end{equation}
where $\|\cdot\|_{2}$ denotes the L2 norm computed over all spatial, temporal, and channel dimensions for a given sample $i$.

\begin{table}[htbp!]
\centering
\caption{Time-averaged relative L2 error on the test set for the PINO-only model versus the full PINO+Diffusion framework. Best performance for each case is in bold.}
\label{tab:l2_errors}
\begin{tabular}{c|c|c}
\hline
\hline
\textbf{Re} & \textbf{PINO-only Error} & \textbf{PINO+Diffusion Error} \\
\hline
100    & \textbf{0.0072}   & 0.0080 \\
250    & 0.0789   & \textbf{0.0213} \\
500    & 0.1676    & \textbf{0.0327} \\
750    & 0.2127    & \textbf{0.0999}    \\
1000   & 0.2548    & \textbf{0.1033} \\
3000   & 0.3271   & \textbf{0.1589} \\
10000  & 0.3914    & \textbf{0.2052} \\
\hline
\hline
\end{tabular}
\end{table}

The results in Table~\ref{tab:l2_errors} quantitatively confirm the behavior observed in the qualitative and spectral analyses. In the laminar flow regime ($\mathrm{Re} \le 250$), the PINO-only model demonstrates high accuracy, and for $\mathrm{Re}=100$, its performance is already excellent, rendering the diffusion-based correction superfluous. The critical role of the diffusion enhancement becomes evident as the system enters the transitional regime at $\mathrm{Re}=750$. At this point, the PINO-only error increases sharply to 21.3\%, marking the onset of its failure to capture the flow's complexity. In stark contrast, the PINO+Diffusion framework contains this error growth, yielding a final error of only 9.9\%.

This trend continues at $\mathrm{Re}=1000$, where the framework reduces the relative error by a factor of approximately 2.5, from 25.5\% to a respectable 10.3\%. This robustly demonstrates the framework's capability to produce accurate surrogate models for systems well into the onset of the turbulent regime, effectively overcoming the spectral limitations of the operator alone. The performance in more intensely turbulent flows is discussed next.

\subsection{Performance at High Reynolds Numbers}

Finally, we assess the framework's performance in the most challenging high-Reynolds-number regimes tested. At Re=3000, the PINO-only error grows to 32.7, while the full framework maintains a respectable error of 15.9. The foundation for this strong quantitative result is the framework's ability to maintain physical fidelity, which is most clearly demonstrated by the energy spectrum in Fig.\ref{fig:spectra_comparison}(b). The framework's prediction closely tracks the ground truth across the low- and mid-wavenumber ranges, indicating a correct distribution of energy across the most significant scales, with only a minor deviation at the highest frequencies. This spectral accuracy is reflected in the temporal snapshots (Fig.\ref{fig:diffusion_re3000}), which show that the PINO+Diffusion model successfully captures the complex evolution of the magnetic vector potential, avoiding the progressive smoothing that plagues the PINO-only approach.

For the most extreme case at $\mathrm{Re}=10000$, while the absolute error increases to 20.5\%, this still represents a halving of the untenable 39.1\% error from the PINO-only model. This substantial and consistent relative improvement demonstrates that the framework continues to capture the essential characteristics of the flow even when point-wise accuracy decreases. As visualizations provided in the Appendix confirm, the full framework retains the large-scale morphology of the complex flow, whereas the PINO-only predictions devolve into unphysically smooth states. This highlights the framework's utility for producing statistically meaningful surrogates in challenging, high-Re scenarios where baseline operators completely fail.

The successful application of this two-stage model presents a promising path forward for surrogate modeling of complex physical systems. By delegating the learning of low-frequency, deterministic evolution to an efficient operator and the high-frequency, stochastic details to a generative model, the adopted framework overcomes a key bottleneck in prior approaches. This methodology of separating and conquering distinct spectral regimes may prove to be a valuable strategy for developing fast and accurate surrogates for a wide range of multi-scale problems in computational physics.

\subsection{Generalization Across Reynolds Numbers}
\label{subsec:generalization}

Our methodology relies on training specialized models for each distinct physical regime. To empirically validate this choice and explore the framework's generalization capabilities, we conducted a series of cross-regime experiments. First, we assessed performance across a large parametric gap by applying our trained $\mathrm{Re}=1000$ model to the $\mathrm{Re}=100$ test set. The results confirmed a fundamental failure in generalization, with the time-averaged relative L2 error exceeding 0.85. This is attributed to the critical mismatch in learned physical priors: the model trained on turbulent statistics attempted to impose unphysical, high-frequency dynamics onto the smooth, laminar flow. This finding provides strong empirical justification for the necessity of regime-specific models when bridging disparate physical systems.

Next, we investigated generalization to a physically similar regime by applying the same $\mathrm{Re}=1000$ model to a dedicated test set of 100 simulations generated at $\mathrm{Re}=900$, following the same methodology described in Section~\ref{subsec:data_generation}. While the PINO-only model still yielded a substantial L2 error of approximately 0.40, demonstrating the sensitivity of the deterministic operator to physical parameters, the full PINO+Diffusion framework proved remarkably robust. The diffusion model, conditioned on the imperfect PINO prediction, successfully corrected for the parametric mismatch, reducing the final L2 error to a promising $\sim$0.13. This result highlights the powerful corrective and adaptive capability of the generative diffusion stage.

These findings collectively suggest that while our framework cannot bridge disparate physical regimes, it exhibits robust potential for interpolation and adaptation between nearby Reynolds numbers. This opens a promising avenue for future work, where a pre-trained model could serve as a powerful foundation, being efficiently adapted to new physical parameters via fine-tuning of the diffusion component, thereby reducing the computational cost of developing new surrogate models.

\section{Summary and Conclusions}
\label{sec:conclusion}

In this work, we addressed the challenge of constructing accurate surrogate models for two-dimensional, incompressible MHD, particularly in regimes characterized by complex, multi-scale turbulence. While PINOs are effective in capturing large-scale, equation-constrained dynamics, we demonstrated that they are fundamentally limited by an inherent spectral bias that suppresses high-wavenumber content~\cite{Rosofsky2023MHD}. This limitation becomes especially pronounced at moderate to high Reynolds numbers, where fine-scale magnetic and velocity structures dominate the system evolution.

To overcome this bottleneck, we evaluated a two-stage framework that couples a deterministic PINO with a conditional, score-based diffusion model. While such hybrid approaches have recently been introduced in other domains~\cite{OommenEtAl2025}, this work presents the first successful application to MHD, yielding state-of-the-art performance across both transitional and turbulent regimes.

The framework's quantitative effectiveness is evident across a wide range of Reynolds numbers. While the PINO model performs well in the laminar regime ($\mathrm{Re}=100$), its error grows substantially in more complex flows. At $\mathrm{Re}=750$, our diffusion-enhanced model cuts the PINO-only error of 21\% by more than half to 9.9\%. This trend continues at $\mathrm{Re}=1000$, where our approach achieves a nearly threefold reduction in error, from 25.5\% to 10.3\%. This improvement becomes crucial in highly turbulent conditions where the PINO-only model exhibits catastrophic failure, generating unphysical, overly smooth states. In stark contrast, our diffusion-enhanced framework maintains physical plausibility, a qualitative success that is mirrored by a dramatic quantitative error reduction: from 32.7\% to 15.9\% at $\mathrm{Re}=3000$, and from 39.1\% to 20.5\% in the most extreme case tested ($\mathrm{Re}=10000$).

In addition to reducing pointwise errors, the framework accurately reconstructs the kinetic and magnetic energy spectra across the entire wavenumber range, demonstrating the model's ability to recover both low- and high-frequency dynamics. Notably, it corrects the spectral deficit in the magnetic field—often the first quantity to exhibit high-frequency degradation in baseline models—restoring spectral consistency even at $\mathrm{Re} = 3000$.

These findings establish a promising paradigm for surrogate modeling in computational physics. By delegating the learning of low-frequency, deterministic evolution to a neural operator and the high-frequency, stochastic corrections to a generative model, the proposed framework provides a principled way to overcome the spectral limitations of existing approaches. Our cross-regime analysis further validates this paradigm, demonstrating not only the necessity of specialized models for disparate physical regimes but also the framework's robustness to small parametric perturbations. This divide-and-conquer strategy is likely to generalize to a broad class of multi-scale problems in fluid dynamics, astrophysics, and plasma physics.

We conclude by emphasizing that our analysis has been confined to the two-dimensional, incompressible, resistive MHD system at $\mathrm{Pr_m} = 1$. This deliberate choice enabled controlled dataset generation and systematic evaluation of the DINOs framework while retaining essential turbulent phenomena such as cascades and reconnection. The extension of this approach to fully three-dimensional turbulence---where anisotropy, Alfvénic interactions, and small-scale intermittency become central---as well as to regimes with magnetic Prandtl numbers significantly different from unity, are active directions of our ongoing research. These forthcoming studies will assess the broader applicability of DINOs across the diverse parameter regimes relevant to plasma physics and astrophysics.

Building upon our generalization study, a key direction for future work is the exploration of fine-tuning pre-trained models as a highly efficient strategy for adapting the framework to new, nearby physical parameters. Future work may also explore infusing physics-based constraints—such as PDE residuals—into the diffusion stage to enhance physical consistency~\cite{Shu2022PID}, as well as extending the framework to kinetic plasma models. These directions offer compelling opportunities to further improve fidelity, generalizability, and interpretability in the data-driven modeling of complex physical systems.

\section*{Acknowledgements}
\noindent EAH acknowledges the partial support of NSF grants
OAC-2514142 and OAC-2209892. 
RH acknowledges partial support from NSF grants OAC-2004879, OAC-2005572, OAC-2103680, OAC-2310548, OAC-2411068.
An award for computer time 
was provided by the U.S. Department of Energy’s (DOE) 
Innovative and Novel Computational Impact on Theory and 
Experiment (INCITE) Program. This research used supporting 
resources at the Argonne and the Oak Ridge Leadership 
Computing Facilities. The Argonne Leadership Computing 
Facility at Argonne National Laboratory is supported 
by the Office of Science of the U.S. DOE under Contract 
No. DE-AC02-06CH11357. The Oak Ridge Leadership 
Computing Facility at the Oak Ridge National Laboratory 
is supported by the Office of Science of the U.S. DOE 
under Contract No. DE-AC05-00OR22725. This research used 
both the DeltaAI advanced computing and data resource, 
which is supported by the National Science Foundation 
(award OAC 2320345) and the State of Illinois, and the 
Delta advanced computing and data resource which is 
supported by the National Science Foundation 
(award OAC 2005572) and the State of Illinois. 
Delta and DeltaAI are joint efforts of the 
University of Illinois Urbana-Champaign and 
its National Center for Supercomputing Applications.

\section*{Code Availability}
\noindent The complete source code for the Diffusion-Integrated Neural Operators (DINOs) framework developed for this study, including all modules for data generation, model training, and analysis, is publicly available in the official repository at \texttt{https://github.com/semihkacmaz/DINOs}. Additionally, the trained models are deposited in the Garden, a platform for sharing and running scientific AI models, and are available at the following address: \texttt{https://thegardens.ai/\#/garden/10.26311\%2Fc4bj-8h61}.

\appendix
\section*{Appendix A: Detailed Results for the High-Turbulence Regime}
\label{app:re10k}

In this section, we provide the detailed visual and spectral results for the most challenging case tested, $\mathrm{Re}=10000$, as referenced in the main body. These figures support the quantitative findings in Table~\ref{tab:l2_errors} and the discussion in Section 3.4.

\begin{figure*}[!htbp]
    \centering
    \includegraphics[width=\textwidth]{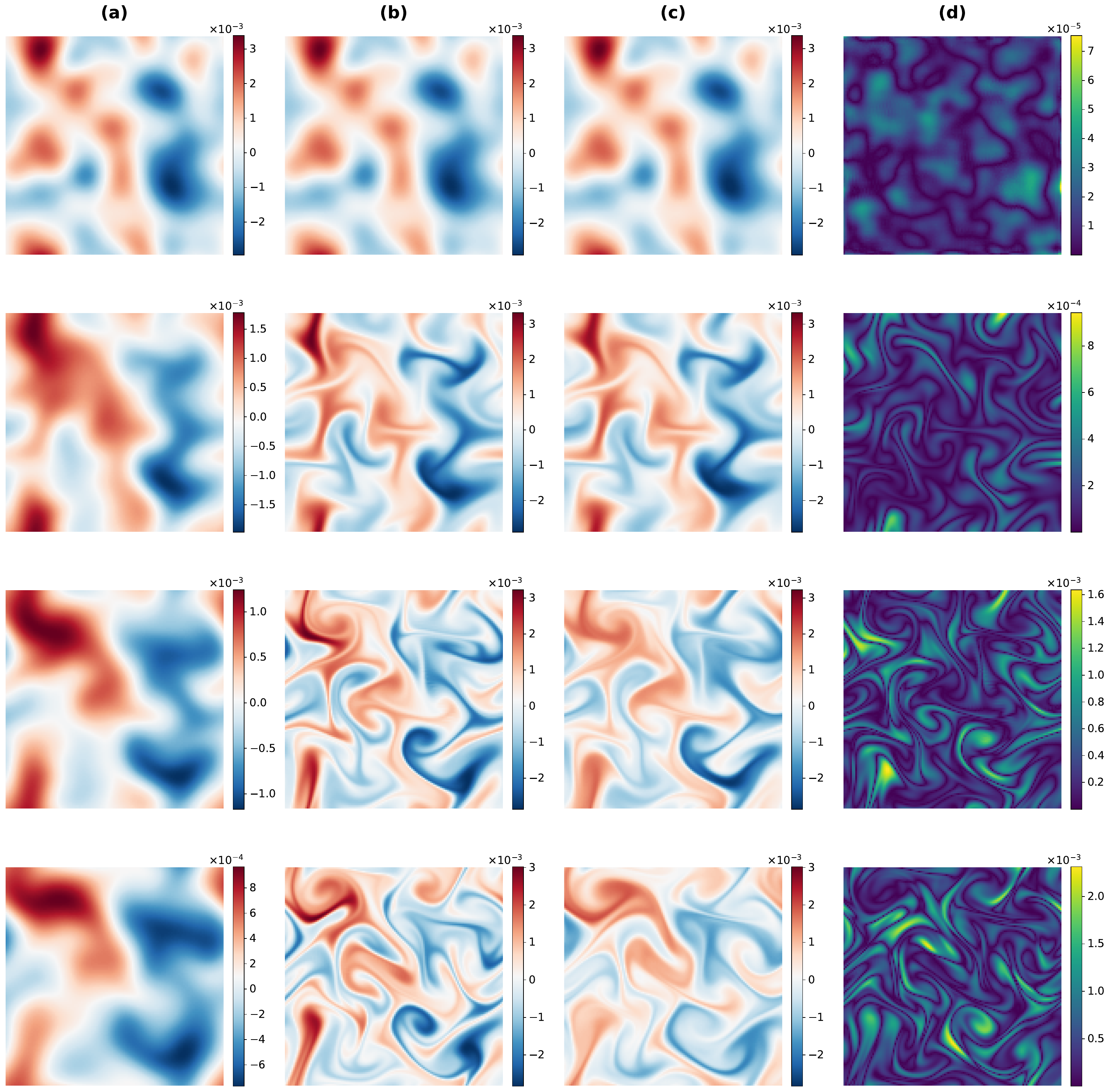}
    \caption{Temporal evolution of the vector potential ($A$) for a turbulent case at $\mathrm{Re}=10000$. The rows display the system at four distinct time steps, corresponding to $t=0, 0.33, 0.67,$ and $1.0$ from top to bottom. Columns show: (a) the smooth PINO-only prediction, (b) the ground truth, (c) the final prediction from the PINO+Diffusion framework, and (d) the absolute error. While point-wise error increases, the framework successfully captures the large-scale morphology of the flow.}
    \label{fig:diffusion_re10000}
\end{figure*}

Figure~\ref{fig:diffusion_re10000} illustrates the temporal evolution of the magnetic vector potential. While the PINO-only prediction (column a) quickly devolves into an unphysically smooth field, the full PINO+Diffusion framework (column c) successfully retains the large-scale morphology of the ground truth flow (column b) throughout the evolution. This demonstrates that even at this high Reynolds number, the framework produces physically plausible results.

\begin{figure}[!htbp]
    \centering
    \includegraphics[width=0.75\textwidth]{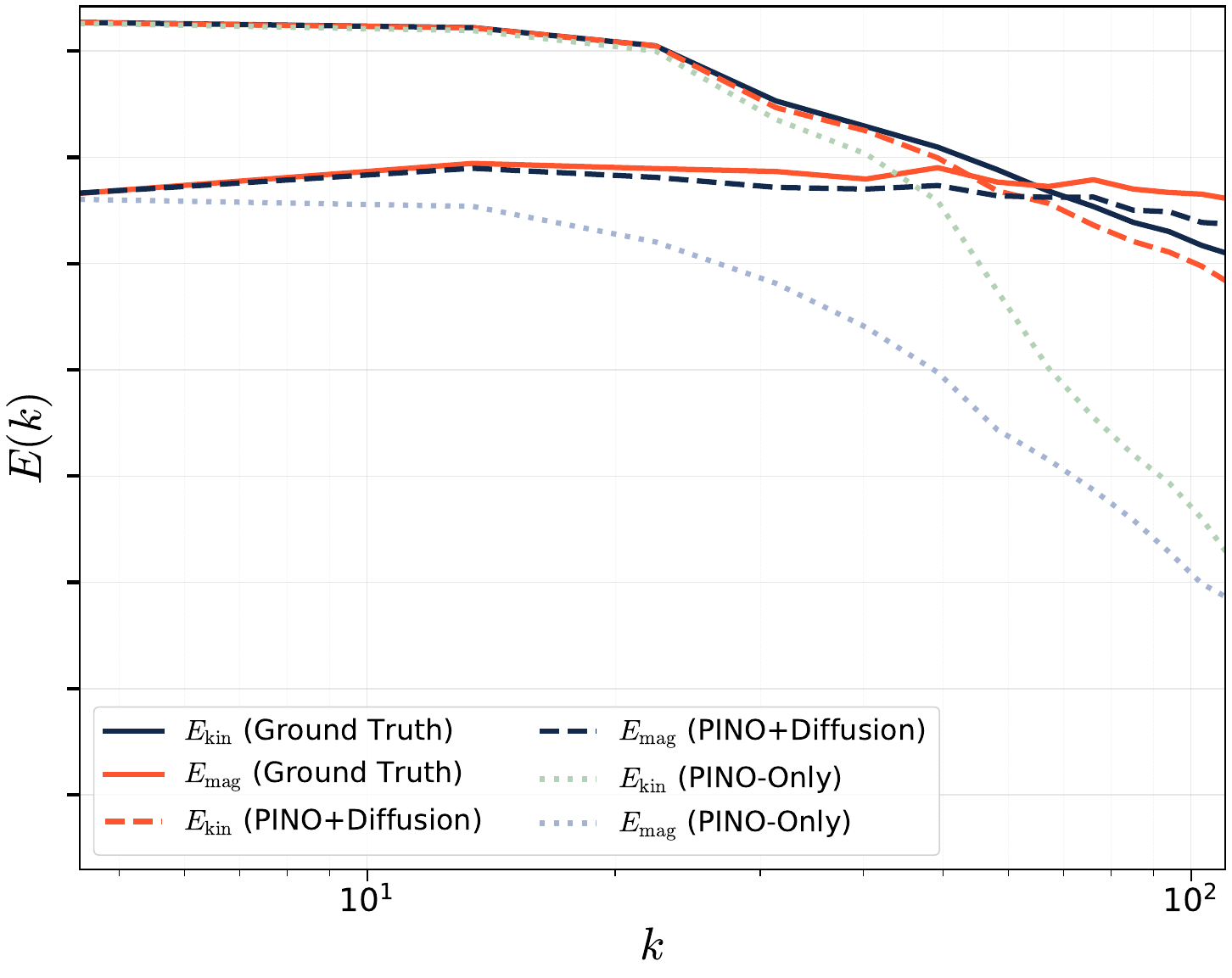}
    \caption{Kinetic and magnetic energy spectra at $t=1.0$ for the $\mathrm{Re}=10000$ case. The PINO+Diffusion model (dashed lines) provides a much-improved spectral reconstruction compared to the PINO-only model (dotted lines), which fails to capture high-frequency energy, particularly in the magnetic field.}
    \label{fig:spectra_re10000}
\end{figure}

This qualitative observation is substantiated by the energy spectra in Fig.~\ref{fig:spectra_re10000}. The analysis confirms that the PINO-only model suffers from a severe spectral bias, failing to represent energy content at high wavenumbers. In contrast, the PINO+Diffusion model provides a significantly better approximation of the true energy spectrum, confirming its ability to correct for the operator's deficiencies even in this highly turbulent regime.

%% If you have bibdatabase file and want bibtex to generate the
%% bibitems, please use
%%
\clearpage

\bibliographystyle{iopart-num}
\bibliography{references}

\end{document}